\documentclass[twocolumn]{revtex4}
\usepackage{amsmath, amsfonts, amssymb, graphicx, color}

\newcommand{\+}{$^{+}$}

\begin{document}

\title{Towards a quantitative theory of tolerance}

\author{Thierry Mora}
\affiliation{Laboratoire de physique de l'\'Ecole normale sup\'erieure,
  CNRS, PSL University, Sorbonne Universit\'e, and Universit\'e de
  Paris, 75005 Paris, France}
\author{Aleksandra M. Walczak}
\affiliation{Laboratoire de physique de l'\'Ecole normale sup\'erieure,
  CNRS, PSL University, Sorbonne Universit\'e, and Universit\'e de
  Paris, 75005 Paris, France}

\begin{abstract}
A cornerstone of the classical view of tolerance is the elimination of
self-reactive T cells during negative selection in the
thymus. However, high-throughput T-cell receptor sequencing data has so far failed
to detect substantial signatures of negative selection in the observed
repertoires. In addition, quantitative estimates as well as
recent experiments suggest that the elimination of self-reactive T cells is
at best incomplete. We discuss several recent theoretical ideas
that can explain tolerance while being consistent with these observations, including collective
decision making through quorum sensing, and sensitivity to change
through dynamic tuning and adaptation. 
We propose that a unified quantitative theory of
tolerance should combine these elements to explain the plasticity of
the immune system and its
robustness to autoimmunity.
\end{abstract}

\maketitle

\section*{Negative selection: all or nothing?}

Thymic selection is an important step in the generation of mature T
cells that can protect us against foreign pathogens while avoiding autoimmunity.
According to the accepted view \cite{Palmer2003}, negative selection eliminates through apoptosis the T cells that
bind too strongly to self-peptides presented on the major
histocompatibility complexes (MHC) antigen-presenting cells in the thymus.
This ensures that T cells cannot trigger autoimmune reactions. The original
evidence for this elimination is based on a transgenic mouse model where males
and females were compared for the survival of their T cells reactive to a
peptide encoded by the male chromosome \cite{vonBoehmer2008,Kisielow1988}.

Recent studies have questioned the role of deletion as the main
mechanism to avert autoimmunity, by detecting through tetramer binding abundant self-reactive T cells
in the blood of healthy humans
\cite{Yu2015}, by finding T cells reactive to tissue-restricted antigens \cite{Legoux2015}, and by observing their proliferation following the ablation of regulatory T cells \cite{Lee2023}.
In fact, turning the concept of negative
selection into a quantitative theory
of the entire repertoire is a challenging task~\cite{Yates2014},
notably because of the physical constraints implied by the large
number of precise decision making processes that it implies. A theory is needed to quantitatively test the plausibility of the processes, possibly enabling the identification of unforeseen mechanisms in case ``the numbers don't add up."   Here we
first outline two quantitative puzzles that have recently been identified, and then discuss possible partial solutions.

\section*{The failure to distinguish negatively selected TCR}
Ex vivo experiments suggest that selection in the thymus is TCR specific and depends
very sharply on the affinity between the TCR and self-peptides
\cite{Daniels2006}. Therefore, we should expect to observe
clear signatures of this selection in the sequence identity of TCR from
different subsets. Such analyses are now possible thanks to the recent
rise of high-throughput repertoire sequencing experiments \cite{Robins2013}.
Specifically, we expect differences between
sequences that passed or failed TCR selection, but also between
peripheral CD4\+ and CD8\+ cells, since those were subjected to
distinct selection forces. If the decision to
eliminate a cell is a deterministic function of its TCR, then negatively selected
cells should have distinguishable TCR sequences from cells that did not pass
negative selection.

Yet a recent analysis based on the repertoire
sequencing of subsets of thymocytes has failed to reveal
features of the TCR that can predict accurately whether a given cell
will pass thymic selection or not~\cite{Camaglia2023}. Briefly, thymocytes were sorted into double positive, activated double positive (as marked by a Nur77 reporter), dying double positive, and apoptotic CD4 and CD8 single positives (SP), as well as SP from the spleen, and their TCR sequenced. Classifiers were then trained on TCRs from these subsets,
using either neural networks or model-free distributions of 3- and 4-mers within the CDR3.
However, these classifiers could not distinguish an apoptotic thymic SP from a mature splenic SP
at the single sequence level.

A limitation of that study is that apoptotic cells were sorted
using an Annexin V marker (a proxy for cell death), and some of these cells could be dying for
other reasons, including during cell manipulation, which would obscure
the signal.
In addition, each of the alpha and beta chains were sequenced in bulk,
and thus analyzed
separately, while affinity is determined by the combination of both chains.
Although the analysis of antigen-specific TCR subsets suggests that
each single chain carries a lot of information about antigen specificity
\cite{Dash2017}, still part of the signal may be lost when considering chain
separately. Another issue is that these differences may be very hard
to detect because of computational reasons. Think of the effect of
negative selection by each self-peptide as digging a ``hole'' in the
potential repertoire. Given the very large number and diversity of
self-peptides, detecting all these holes may be too hard for our
computational techniques. Structural data and experiments \cite{Stadinski2016} suggest that the TCR makes contact with the peptide at only a few sites, where hydrophobic residues are suppressed by negative selection, consistent with theoretical predictions \cite{Kosmrlj2008}, but it is not clear how to identify those sites.
Despite these caveats, which may smear and reduce the 
signal, the poor discriminability of passing versus failing sequences
is still puzzling.

We should also be able to detect different selection pressures
on the TCR of CD4\+ and CD8\+ cells, since they interact with
the two distinct MHC-II and MHC-I classes,
and hence with separate sets of self-peptides during negative
selection. These differences turn out
to be statistically significant at the repertoire level
\cite{Emerson2013},
but too small to allow discriminating
CD4\+ or CD8\+ cells with good accuracy based on their
TCR only~\cite{Isacchini2021,Camaglia2023}. Unlike the
discrimination between dying and surviving cells, CD4\+ and CD8\+
repertoires need not be exclusive: in principle the same TCR
could be found in both repertoires, and this could in part explain why
such a discrimination task cannot be perfect.

By the same logic, we should also be able to distinguish the TCR of regulatory T cells (Treg).
Tregs are CD4\+  T cells that may recognize the same antigens as conventional T cells (Tconv),
but with a bias for self-antigens, and
inhibit the inflammatory response
caused by Tconv activation. For this reason Tconvs and Tregs may share the same receptor. 
TCR statistical scores have been learned from repertoire data that can identify
modest but measurable differences between the
Treg and Tconv populations~\cite{Isacchini2021,Lagattuta2022}. Again, they are not powerful
enough to classify individual TCRs. A similar observation was
made when trying to predict self-reactivity (measured by CD5 level) from the TCR
sequence \cite{Textor2022}.

In summary, while our idea of how thymic selection works suggests
there should be strong signatures of fate at the sequence level, in
practice it seems difficult to identify sequence signatures of TCR
that determines unambiguously their ability to survive thymic selection or to
fall into a given subpopulation.

\section*{Can T cells screen all self-peptides?}
The second puzzle, first noted by Butler et al.~\cite{Butler2013a}, concerns the numbers and timescales involved in
the negative selection process. In principle, to avoid autoimmunity each T cell should be
screened against every presentable self peptide-MHC complex. Do
they have enough time to do that?
Each T cell spends about 4-5 days in the thymus,
which gives them time to interact with about 500 antigen presenting
cells (APC)~\cite{LeBorgne2009}. During each of those encounters, 
multiple copies of the TCR on the cell surface may bind to
distinct peptide-MHC complexes presented by the APC. In principle, 
any of those engagements could result in T-cell activation and
subsequent apoptosis. This means
that the number of self peptides that may be screened during each 
encounter with an APC during negative selection could be large.

Unfortunately, it is difficult to estimate that number.
Previous literature put the total number
of screened peptides (across APC encounters) in the thousands \cite{Faro2004}.
Butler et al.~\cite{Butler2013a} used a self-consistency argument to
estimate that number, by quantifying its impact on the specificity of selected
TCR: their model predicts that the fraction of peptides recognized by a peripheral TCR is
inversely related to the number of screened self-peptides during
negative selection. Using  peptide-specific precursor frequencies
measured in mice~\cite{Jenkins2012}, they estimate the number of
screened peptides to be $\sim 7,000$, or $\sim 2-10\%$
of the self-peptidome (20,000 mamalian genes each having 300 peptides, and assuming that 1-5\% of peptides are presented by MHC). These numbers suggest T cells may not have enough time to make sure they are not self-reactive.

\section*{The unlearnability of the self peptidome}
This issue of insufficient screening time could be solved if the
space of self-peptides could be ``learned,'' i.e. if there were general rules or properties that distinguish self- from pathogen-derived peptides.
TCR could then generalize their knowledge from their interaction with a few self-peptides and would not have to scan every one of them. This idea of learnability was proposed in Refs.~\cite{Wortel2020,Mayer2022}. However, both studies concluded that while there are minute statistical differences between self and pathogen-derived peptides, they are too small to be used for efficient self vs nonself discrimination. In other words, the difference been self and non-self peptides cannot be learned through rules, and TCR actually do need to scan them exhaustively, i.e. learn them ``by heart'' (akin to overfitting in machine learning language).

This conclusion is consistent with observations that a single mutation in an epitope can turn an non-immunogenic self-peptide into an immunogenic neoantigen \cite{Matsushita2012}, and more generally that self- and pathogen-derived antigens from databases such as IEDB \cite{Vita2015} are promiscuous in sequence space \cite{Mayer2022}. Self-peptide learnability also seems implausible from an evolutionary perspective. If rules existed to recognize self-peptides,
pathogens would be under strong selective pressure for their peptidome to mimic those rules. Finally, self-peptidome learnability would still imply discriminability of self-reactive TCR, and thus cannot explain why we fail to observe it when analyzing their repertoires.

\section*{Quorum sensing}
A second solution, which can explain both our puzzles, was proposed by Butler et al.~\cite{Butler2013a}. It relies on the idea (similar to the Condorcet Jury theorem) that if $N$ cells have to make a decision, and the probability that each cell makes the right one is better than chance, then the probability that a collective decision by majority vote gets it right---or equivalently that a ``quorum'' of activated cells is reached \cite{Burroughs2006a}---quickly goes to 1 as $N$ increases.
Experiments suggest that T-cell decision making depends on the number of activated cells, through the secretion and sensing of cytokines such as Interleukin-2 \cite{Feinerman2010,Polonsky2018}. What distinguishes a self from a foreign peptide is that at least {\em some} of the self-reactive TCR are removed, meaning that the number of precursor T cells specific to self-peptides falls below the quorum necessary to start an immune reaction, while the number of precursor T cells specific to foreign peptides rises above. In Ref.~\cite{Butler2013a}, the optimal quorum is estimated to be around 40 T-cells, separating the case of self-peptides recognized by $10$-$30$ T cells, from foreign peptides, which would recruit $50$-$100$ cells. Recent experiments have since provided evidence that T-cell fate is indeed collective and requires a similar quorum of activated cells \cite{Polonsky2018,Zenke2020}.

This argument relies on the assumption that the pre-selection frequency of peptide-specific precursors is relative constant across peptides. However, this frequency is known to vary by one or two orders of magnitude, even across foreign epitopes \cite{Moon2007,Legoux2010,Jenkins2012}, and may be predictive of immune response magnitude and immunodominance \cite{Moon2007,Jenkins2012}. Imagine a self-peptide with pre-selection frequency of 200 precursor T cells, half of which are removed by negative selection. From the quorum perspective, this is indistinguishable from a foreign peptide with a pre-selection frequency of 100 T cells, all of which survive selection. Thus, in order to avoid auto-immunity, it is important that the absolute number of self-peptide-specific T cells be controled, and not just their probability of removal. This implies the existence of other mechanisms of regulation.

\section*{Regulation, dynamic tuning, and adaptation}
The third solution that we want to discuss here relies on the immune system adjusting its reponse through dynamic tuning \cite{Grossman2001,Bhandoola2002}. In its simplest form, immune perturbations activate both excitatory and inhibitory pathways. These two pathways may either compensate each other, resulting in adaptation and tolerance, or trigger an immune response if the excitatory signal is dominant. Dynamic tuning is reminiscent of mechanisms of adaptation discussed in the context of simple signaling networks, such as the chemotactic network of the flagellar bacterium {\em E. coli} \cite{Tu2008}. In these systems, the activation of the pathway on short time scales is accompanied by repressive action on longer times, resulting in transient activation followed by return to a low level of activity, even if the perturbation is sustained.
A similar idea has been theorized in the ``discontinuity theory'' of immunity \cite{Pradeu2013}, mostly discussed in the context of the innate immune system. According to this theory, the state of the immune system does not depend on immune stimuli {\em per se}, as long as these stimuli do not change over time. Rather, it responds to the kinetics of the antigenic or immune environment \cite{Johansen2008}, allowing the system to detect perturbations and to adapt to sustained changes.

For T cells, obvious candidates for repressive regulatory signals are Tregs and anergy \cite{Kalekar2017,Alonso2018}. Anergy is a state of T cells where their ability to proliferate is impaired. This can happen when cells are stimulated weakly or in absence of co-stimulatory signals, as happens when engaging self-peptides in absence of inflammation. Regulatory Tregs are selected in the thymus to be more self-reactive Tconvs, and are essential to prevent autoimmune diseases.
They proliferate in response to an immune challenge, and tend to suppress the immune response through cytokine signaling \cite{Busse2010,Feinerman2010}, by inducing anergy \cite{Alonso2018}, or by pruning self-activated T cells \cite{Wong2021}.

This double mechanism of activation by effector T cells, and repression by regulatory T cells, is very much reminiscent of an incoherent feedforward loop. This regulatory architecture causes the system to respond to fold-changes rather than to the absolute value of the perturbation \cite{Goentoro2009}, and allows for a quick but controled response \cite{Rosenfeld2002}. It is consistent with the ideas of dynamic tuning and discontinuity theory. The same tug-of-war phenomenon between effector and regulatory T-cell is central to the mechanism of quorum sensing discussed earlier \cite{Burroughs2006a, Feinerman2010}.

Marshland et al~\cite{Marsland2021} proposed a theory of the balance between effector and regulatory T cells in the stationary state (constant antigenic environment), based on an ecological description of interactions between T cells and antigens. In their model, self-peptides stimulate Tconvs and Tregs in the same way such that the number of Tconvs and Tregs always
counterbalance each other. They show that a minimum number of distinct Treg specificities is required to satisfy this balance and thus to avoid autoimmunity.

In these theories, the fact that Tregs are more self-reactive  than Tconvs provides a natural mechanism for selectively suppressing autoimmunity, regardless of negative selection. Self-reactive TCR are acceptable as long as they are not
  massively stimulated \cite{Zehn2006}. Accordingly, self-tolerance may be disrupted when
  a self-antigen or a cross-reactive foreign antigen is overexpressed.

It is not clear how to reconcile these theories of adaptation with the immunology
of asymptomatic chronic infections such as Cytomegalovirus, in which the antigen stimulation is stable, yet mobilizes a large growing fraction of the TCR repertoire \cite{Schober2018}. It could be explained with a high adaptation plateau where the response is repressed but
this remains to be checked quantitatively. Another caveat of these theories is that Tregs make up a small fraction of the repertoire, about 5-10\% of CD4\+ T cells. Whether these small numbers are sufficient to regulate and repress autoimmunity is a quantitative question that would require further investigation.

\section*{Concluding remarks}
We presented quantitative observations that seem to challenge the classical view of negative thymic selection, which holds that self-reactive T cells should be largely eliminated from the periphery. Since these observations are based on indirect estimates and incomplete data, it is possible that these contradictions will evaporate upon more rigorous inspection (see Outstanding Questions box). In the modern view of tolerance, autoimmunity is averted by a combination of mechanisms, among which negative selection is an important but not sufficient element. A future quantitative theory will have to
combine a few of the solutions outlined above, in particular the ideas of quorum sensing and adaptation, with the hope that these ideas can mutually correct each other's inconsistencies. This theory should aim to explain existing data and to make new testable predictions.

\section*{Acknowledgements}
The authors thank Gr\'egoire Altan-Bonnet, Benny Chain, Arup Chakraborty, and Olivier Lantz for
useful discussions. This research was supported by the European Research Council COG
724208 and ANR-19-CE45-0018 ``RESP-REP'' from the Agence Nationale de la Recherche.

\bibliographystyle{pnas}

\end{document}